# The Trusted Caregiver: The Influence of Eye and Mouth Design Incorporating the Baby Schema Effect in Virtual Humanoid Agents on Older Adults Users' Perception of Trustworthiness

JENNIFER HU*, School of Art and Archaeology, Zhejiang University, Hangzhou, China

The increasing proportion of the older adult population has made the smart home care industry one of the critical markets for virtual human-like agents. It is crucial to effectively promote a trustworthy human-computer partnership with older adults, enhancing service acceptance and effectiveness. However, few studies have focused on the facial features of the agents themselves, where the "baby schema" effect plays a vital role in enhancing trustworthiness. The eyes and mouth, in particular, attract most of the audience's attention and are especially significant. This study explores the impact of eye and mouth design on users' perception of trustworthiness. Specifically, a virtual humanoid agents model was developed, and based on this, 729 virtual facial images of children were designed. Participants (N=162) were asked to evaluate the impact of variations in the size and positioning of the eyes and mouth regions on the perceived credibility of these virtual agents. The results revealed that when the facial aspect ratio (width and height denoted as W and H, respectively) aligned with the "baby schema" effect (eye size at 0.25W, mouth size at 0.27W, eye height at 0.64H, eye distance at 0.43W, mouth height at 0.74H, and smile arc at 0.043H), the virtual agents achieved the highest facial credibility. This study proposes a design paradigm for the main facial features of virtual humanoid agents, which can increase the trust of older adults during interactions and significantly contribute to the research on the trustworthiness of virtual humanoid agents.

---

* corresponding author

# 1 INTRODUCTION

With advancements in technology and the growing demand for emotional support, AI-driven virtual agents have shown immense potential in fields such as education (Fitton, 2020), entertainment (Torre, 2019), and therapeutic assistance (Catania, 2023). At the same time, developments in deep learning and neural networks have endowed virtual agents with the advantages of being lightweight, customizable, and highly flexible (Ivanovic, 2022), expanding their application value across various fields. This adaptability allows them to meet the needs of diverse scenarios, providing innovative interaction experiences for various industries. As the aging population grows, smart home care has become an important application area for virtual agents (Appel, 2020). Virtual agents not only provide emotional support for older adults but also assist in enhancing life skills and social participation, significantly improving quality of life and bridging the digital divide (Wang, 2022), meeting the technological needs and expectations of older adults. However, virtual agents face challenges, such as insufficient emotional empathy, privacy and security risks, and reliability issues during long-term interactions, especially when addressing complex interpersonal needs (Bickmore & Picard, 2005). Trust is particularly crucial in older adults living environments because it directly affects their acceptance and reliance on virtual agents (Kim, 2020). Virtual agents' appearance and behavior influence trust and are closely tied to their reliability and consistency during interactions.

Previous studies have shown that the more human-like a virtual agent's appearance, the stronger the user's sense of trust (Sun, 2021). Humans are also more likely to be attracted to anthropomorphized features (Landwehr et al., 2011). The facial image of a virtual agent forms the first impression in users' human-computer social cognition, which is critical for building trust and enhancing overall experience (Clark, 2020), especially when older adults are adopting assistive technologies (Esposito, 2019). Interestingly, research has found that baby-like faces in digital interfaces are more likely to win trust than adult or older adults faces (Aljukhadar, 2010). This "baby schema" effect, which refers to faces with infant-like characteristics, significantly increases the credibility of virtual agents (Navarini, 2021; Luo, 2020). Previous research has focused on the impact of facial features in virtual agents on trust, mainly how features like the eyes, mouth, face roundness, and forehead size in baby-like faces contribute to perceptions of honesty (Masip et al., 2004). Additionally, some studies have explored how abstract virtual characters with fewer facial features influence user perception (Ferstl et al., 2020). However, these studies often overlook the interdependence between features or the effects of extreme shifts in feature placement and lack consideration for cross-cultural and diverse designs for particular groups, such as older adults. When judging facial credibility, older adults may be influenced by cognitive aging, emotional needs, life experience, and social support dependence. Moreover, research has shown that eyes and mouth, as critical areas for emotional expression, play a significant role in how older adults judge facial trustworthiness (Adolphs, R., 2002). Therefore, how older adults uniquely assess the facial credibility of virtual agents remains an area for further investigation (Dias, 2023; Rheu, 2021). This study explores the impact of critical facial features on older adults' perception of trustworthiness in virtual agents. Based on feature placement and shifts, the study will focus on the size and position of the eyes and mouth to optimize the facial features of virtual agents. We hypothesize that these two features play a critical role in conveying trustworthiness. The experiment will manipulate and combine the size and position of the eyes and mouth in virtual agents' faces to test whether the "baby schema" effect can be effectively applied to virtual agent design for older adults. The findings will provide valuable insights into the impact of facial features on trustworthiness and offer guidance for optimizing virtual digital humans, improving the well-being and user experience of the older adult population.

## 2 RELATED WORK

### 2.1 Facial features of virtual humanoid agents

The study of facial features in virtual humanoid agents has long been an intriguing and prominent research focus within human-computer interaction and computer graphics (Shi, 2020). These virtual agents possess complex facial characteristics, including expressions, eyes, mouths, and other crucial details. These directly and profoundly impact users' perception, emotional communication, and cognitive experience during interactions with virtual characters (de, 2015). A critical study by Ylva Ferstl examined the influence of facial width and eye size on personality perception. By carefully designing virtual characters with realistic feature dimensions, this study provided valuable insights into character design for video games, movies, and specific applications involving virtual agents (Ferstl, 2017). Understanding how facial features affect users' perception and connection with virtual beings holds immense potential for creating more relatable and engaging virtual interactions.

Research in this area has also shown that consistency between virtual human facial expressions and emotional content positively influences user acceptance and comprehension of information (Mudrick, 2017). When virtual agents can convey emotions and expressions authentically, users are more likely to connect with them on a deeper level, enabling smoother and more meaningful interactions. Arturo S. and colleagues designed a series of virtual human facial expressions for emotional communication experiments, confirming that humans could accurately recognize emotions in virtual faces with an accuracy rate close to 90% (Garcia, 2020). Notably, neutral, happy, and angry expressions were the easiest to recognize (Garcia, 2020). Despite significant progress in this field, there remains to be more room for further research and development. The ongoing aim is to enhance the realism, expressiveness, emotional conveyance, and perception of virtual human facial features. As technology advances, unlocking the full potential of virtual agents to evoke authentic emotional responses and foster more immersive experiences has become an exciting and continually evolving pursuit.

### 2.2 The "baby schema" effect

The "baby schema" effect (Lorenz, 1943) describes a set of characteristic infant features, such as a round face, large head, big eyes, high forehead, chubby cheeks, small nose, and small mouth (Prokop, 2018). Faces exhibiting a robust "baby schema" effect also tend to display some typically feminine features (Koyama, 2006) and bear a degree of resemblance to female faces (Hess, 2009). Interestingly, this sensitivity to "baby schema" is not exclusive to infants; it also influences the perception of baby-like features in adult faces (Kuraguchi, 2015). These infant-specific features often evoke positive emotions in observers (Lehmann, 2013; Luo, 2020). Studies have shown that seeing "baby schema" characteristics leads to more favorable attitudes and emotional responses, as evidenced by changes in both peripheral and central nervous system activity (Venturoso, 2019). Other studies have shown that the "baby schema" effect can elicit positive emotional reactions among older adults during initial impressions (Fukuoka, 2023). Facial features associated with the "baby schema" are vital in attracting attention and conveying credibility. Studies indicate that facial proportions, such as upper face length (forehead length), interpupillary distance (eye distance), and lower face length (chin length), are significant factors in perceived facial attractiveness.

Furthermore, specific features like pupils (Navarini, 2021), expressions (Luo, 2022), and eyebrows (Song, 2021) have been extensively studied concerning perceived trustworthiness. In particular, perceived trustworthiness is primarily influenced by the size and position of the eyes and mouth. Although the nose occupies a substantial portion of the face, empirical evidence suggests that it plays a lesser role in conveying anthropomorphic trustworthiness than the eye and mouth areas (Song, 2021). The roundness and largeness of baby-like eyes, typical of infants under one year, have been associated with high credibility (Ferstl, 2017). The light sclera seen in infants enhances gaze direction discrimination, allowing observers to infer intent (Kleisner, 2010). However, prolonged eye contact with large eyes can also evoke feelings of seriousness or even fear (Kleberg, 2021). Similarly, in the context of mouth size, research on male faces has shown that a wider mouth enhances perceptions of dominance and leadership ability (Re, 2016). Positional shifts in facial features also play a role: human infant eyes are positioned relatively closer together and lower on the face than in adults. Infant facial features generally exhibit an inward orientation, with babies displaying higher foreheads (Miesler, 2011), closer-set pupils (Venturoso, 2019), and shorter chins (Montepare, 2003).

### 2.3 Credibility evaluation

Due to the higher attractiveness and lower conflict risks associated with physically attractive young females, as observed in studies on social and evolutionary psychology (Grammer, Karl, et al., 2003; Thornhill, R., & Gangestad, S. W., 1999). People frequently rely on stereotypes when making rapid judgments about others' trustworthiness. In everyday life, facial cues—especially those from the eyes and mouth—serve as essential heuristic indicators that guide these decision-making processes (Todd, 2000). Remarkably, individuals can form credibility assessments based on first impressions in as little as 34 milliseconds, and these impressions remain strikingly consistent even when exposure time exceeds 200 milliseconds (Todorov, 2015). Furthermore, humans can adapt their social judgment strategies when assessing unfamiliar individuals, with gender features influencing these judgments. According to the complementary needs theory, self-similarity only affects the social judgments of same-sex individuals in real-life situations. In contrast, individuals prefer those who differ from themselves when evaluating the opposite sex (Nakano, 2022).

The features associated with the "baby schema" may be effective predictive indicators of trustworthiness in virtual agents, potentially influencing human intentions to approach and interact with them. Research suggests that trust judgments begin to form in early childhood, with trust impressions developing from age three and reaching maturity around 10-13 years old (Siddique, 2022). However, unlike younger individuals, older adults may exhibit age-related biases in their facial trust judgments due to natural declines in physical and cognitive abilities (Altun, 2022). Therefore, whether the "baby schema" effect can still evoke high levels of trustworthiness in older adults remains a question that warrants further investigation (Fukuoka, 2023).

### 3 MATERIAL AND METHOD

### 3.1 Stimuli

Due to the higher attractiveness and lower conflict risks associated with physically attractive young females (Grammer, Karl, et al, 2003; Thornhill, R., & Gangestad, S. W. (1999). ) this study selected the facial outlines of 4-6-year-old female children (Lizhu Luo, 2020) and adjusted their facial proportions and features according to the characteristics suggested by the "baby schema" effect. The images used by the virtual agent designers were created using the Character Creator platform (Garcia, 2020), following this procedure:

First, a schematic diagram based on the "baby schema" was created, with a rectangle drawn around the facial region containing dense features. The upper boundary of the rectangle was aligned with the horizontal line of the eyebrows, and the lower boundary was aligned with the horizontal line of the lower lip, enclosing the region of interest. Second, horizontal and vertical reference lines were drawn within this enclosed region to mark the positions of the eyebrows, eyes, nose, and mouth. Third, the facial features of the default child model were adjusted according to these reference lines to match the facial characteristics of the "baby schema." This model is called the "base" model, and two operations were performed. First, the facial width-to-height ratio (fWHR) was calculated. The fWHR is derived from the maximum horizontal distance between the cheeks (referred to as cheek width) and the vertical distance from the upper lip to the highest point of the eyelid (Yao, 2022; Clark, 2020). The fWHR for the standard "baby schema" (Figure 1) was calculated as 1.96, while the fWHR for the base model was 1.90, resulting in a relative error ($\Delta$) of 3.00%. This indicates that the base model meets the "baby schema" regarding its facial width-to-height ratio. Second, facial feature assessments were conducted by measuring the distances between selected facial landmarks: forehead length (AO), left eye width (E1F1), correct eye width (E2F2), average eye width (EF, calculated as the mean of E1F1 and E2F2), face width (CD), face length (AB), mouth length (KL), nose length (HG), and nose width (IJ). Based on these measurements, five facial parameters were calculated to assess the "baby schema": AO/AB, EF/CD, HG/AB, IJ/CD, and KL/CD. The calculations for the base model (Figure 2) resulted in AO/AB = 0.19, EF/CD = 0.35, HG/AB = 0.24, and IJ/CD = 0.34, similar to the standard "baby schema" values. Fourth, using a controlled variable approach, multiple copies of the base model were created and divided into six groups, each corresponding to a specific feature change relative to the base model (Group 1: eye size, Group 2: eye height, Group 3: eye distance, Group 4: mouth size, Group 5: mouth height, Group 6: smile arc). The variations in the corresponding features formed a systematic staircase for the experiment. Each group contained three variations of a specific feature. For example, in Group 1 (eye size), the variations included large eyes (0.25W), medium eyes (0.23W), and small eyes (0.21W). Aside from changes in the specific feature being tested within each group, the rest of the facial features remained consistent. Other features, such as facial expressions, body shape, posture, and background, were unchanged (Figure 3). As a result, a total of 729 virtual agent facial images with different feature variations were generated (3 * 3 * 3 * 3 * 3 * 3 = 729).

Therefore, this study aims to expand the theory of facial credibility by systematically manipulating the size and positional displacement of the eyes and mouth and proposes the following hypotheses:

H1: Virtual agents with eye size of 0.25W are perceived as the most trustworthy.

H2: Virtual agents with mouth size of 0.27W are perceived as more trustworthy.

H3: Virtual agents with eye height of 0.62H are perceived as more trustworthy.

H4: Virtual agents with eye spacing of 0.41W are perceived as more trustworthy.

H5: Virtual agents with mouth height of 0.77H are perceived as more trustworthy.

H6: Virtual agents with a smile magnitude of 0.043H are perceived as more trustworthy.

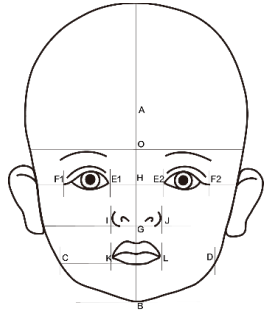

Figure 1: Diagram of measurement parameters for infant facial proportion.

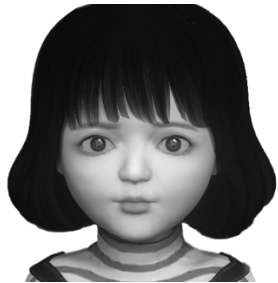

Figure 2: Basic facial design based on proportional parameters.

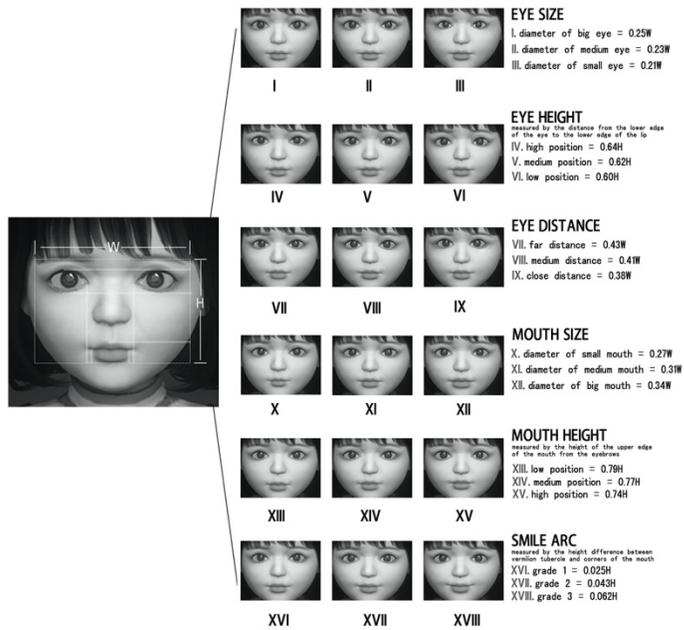

Figure 3: Detailed facial feature indicators for virtual agents. (Due to the three-dimensionality of the face, some values may not show linear changes, but the effect observed from a frontal view is consistent.)

## 3.2 Experimental Design and Participants

To test these hypotheses, we designed a mixed experiment with the size of the eyes and mouth as between-subject variables, the vertical and horizontal positions of the eyes, the vertical position of the mouth, and the vertical position difference between the mouth corners and the philtrum as within-subject variables. One hundred sixty-two participants were recruited from the community to participate in the study. Each participant completed the experiment in a controlled, laboratory-like environment, free from interference by experimenters or other participants (Figure 4a). The experimental procedures were carried out with the assistance of research assistants, and the data were collected via a specially designed website (Figure 4b). The average age of the sample was 69.31 years (SD = 7.902). Detailed demographic information is provided in Table 1. All participants provided informed consent before they participated in the study.

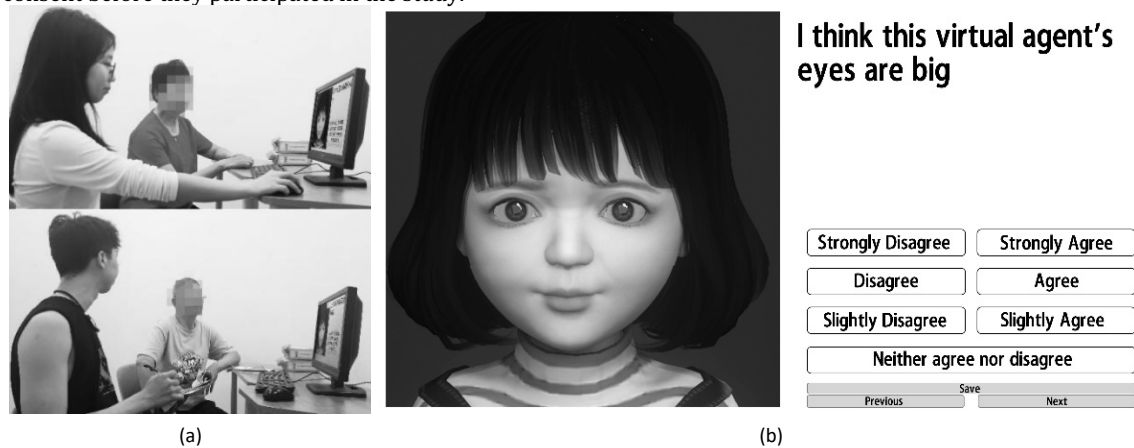

(a) (b)

Figure 4: Scene of the experiment site (a) and User experiment interface (b).

Table 1: Demographic information of the samples in this study.

| Index | Descriptions | Frequency | Percentage |
| --- | --- | --- | --- |
| **Age** | 60-64 | 60 | 37.0% |
| | 65-69 | 39 | 24.1% |
| | 70-74 | 24 | 14.8% |
| | 75+ | 39 | 24.1% |
| **Gender** | Male | 78 | 48.1% |
| | Female | 84 | 51.9% |
| **Educational Level** | Illiteracy | 15 | 9.3% |
| | Elementary school | 69 | 42.6% |
| | Junior high school | 60 | 37% |
| | Technical secondary school and senior high school | 6 | 3.7% |
| | Above senior high school | 12 | 7.4% |
| **Mobile phone experience** | Never used before | 57 | 35.2% |
| | 0-1 year use (exclusive for 1 year) | 6 | 3.7% |
| | 1-2 years use (exclusive for 2 years) | 0 | 0% |
| | More than 2 years use | 99 | 61.1% |

### 3.3 Measurement

Participants were asked to rate their level of agreement with six statements on a seven-point Likert scale (1 = strongly disagree, 7 = strongly agree) after viewing a specific virtual agent design: (1) I think the girl's eyes are a bit too large, (2) I think the girl's mouth is a bit too large, (3) I think the girl's eyes are positioned a bit too high, (4) I think the girl's eyes are spaced a bit too far apart, (5) I think the girl's mouth is positioned a bit too low, and (6) I think the girl is smiling at me. To assess facial credibility, participants were also asked to rate their agreement with five statements on a seven-point Likert scale (1 = strongly disagree, 7 = strongly agree): (1) I think the girl looks reliable, (2) I think the girl looks sincere, (3) I think the girl looks honest, (4) I think the girl looks trustworthy, and (5) I think the girl looks persuasive (Gorn, 2008). Given the internet experience and comprehension abilities of older adult participants, the term "girl's image" was used to describe the appearance of the virtual agent in the digital interface.

### 3.4 Experimental Procedure

After briefly introducing the research background, participants were given an informed consent form and an information sheet. The experiment assistant explained the purpose of the documents and helped participants complete them. The experiment involved six facial features, each with three variations, resulting in 18 independent feature variations (6 features * 3 parameters). This led to 729 possible facial combinations (3 * 3 * 3 * 3 * 3 * 3). Each of the 162 participants was shown a random subset of nine face combinations, with two participants assigned to each subset, resulting in 81 unique subsets. Each group's sequence of virtual agent facial images was randomized to control for within-subject learning effects (Bosmans, 2005). The experimental stimuli were presented through a web interface, and for each stimulus, participants were asked to focus on the facial features of the virtual agent before completing a questionnaire and interview. After finishing the survey, participants were informed that the experiment had concluded and appropriate compensation would be provided.

## 4 RESULT

To examine the hypothesis regarding the influence of the "baby schema" on facial credibility, we conducted descriptive analysis, manipulation check, and a six-way mixed analysis of variance (ANOVA) using SPSS and Stata to examine the effects of facial feature size and position.

### 4.1 Descriptive Analysis and Manipulation Check

To assess the normality of the univariate data, we conducted kurtosis and skewness tests for the six items. Then, descriptive analysis was performed on the different factors in the study. Regarding eye size, the virtual agent faces with more enormous eyes received higher scores (mean = 5.15 vs. 5.20 vs. 5.23; $F(2,1455) = 14.32$, $p < 0.01$). Similarly, for eye height, faces with higher eye positions received higher scores (mean = 5.16 vs. 5.19 vs. 5.22; $F(2,1455) = 14.32$, $p < 0.01$). When it came to the distance between the eyes, faces with greater eye spacing scored the highest (mean = 5.07 vs. 5.12 vs. 5.38; $F(2,1455) = 24.71$, $p < 0.01$).

In terms of mouth position, smiles with upwardly angled mouth corners were perceived more positively (mean = 5.06 vs. 5.19 vs. 5.32; $F(2,1455) = 48.68$, $p < 0.01$). Additionally, smaller mouth sizes were associated with higher scores ($F(2,1455) = 11.92$, $p < 0.01$). For perceived mouth position, lower mouth positions were rated as being more downward (mean = 5.49 vs. 5.03 vs. 5.05; $F(2,1455) = 11.92$, $p < 0.01$). However, mouths in a neutral position were perceived as lower in vertical position than mouths with a "low" position. A summary of the descriptive analysis of these factors is presented in Table 2.

### 4.2 The influence of feature size

#### 4.2.1 Main Effects of Feature Size

In this study, the facial personification credibility questionnaire (introduced in 3.3 measurements, five items on a seven-point Likert scale) was used to measure the level of facial trustworthiness among older adults, and the items showed high internal consistency (Cronbach is 0.954). After testing for homogeneity of variances and regression coefficients, we conducted covariance analyses (ANCOVA) to examine the effects of eye and mouth size (between-subject factors) and eye and mouth position (within-subject factors), with prior smartphone usage experience and gender as covariates. Table 3 summarizes the results of the ANCOVA. We focused primarily on the main effects of feature size and position, as they are theoretically relevant to the research question.

Table 2: Summary of Descriptive Analysis of Different Factors.

| Factors | Levels | Mean | SE | 95% Confidence Interval | |
|---|---|---|---|---|---|
| | | | | Lower Bound | Upper Bound |
| Eye size | Small | 5.15 | 0.96 | 3.27 | 7.03 |
| | Medium | 5.20 | 0.96 | 3.30 | 7.09 |
| | Big | 5.23 | 0.98 | 3.31 | 7.14 |
| Mouth size | Small | 5.49 | 0.86 | 3.79 | 7.18 |
| | Medium | 5.03 | 1.01 | 3.05 | 7.01 |
| | Big | 5.05 | 0.95 | 3.18 | 6.93 |
| Eye height | Low | 5.16 | 0.98 | 3.24 | 7.08 |
| | Medium | 5.19 | 0.96 | 3.31 | 7.08 |
| | High | 5.22 | 0.96 | 3.33 | 7.11 |
| Eye distance | Close | 5.07 | 1.07 | 2.97 | 7.17 |
| | Medium | 5.12 | 0.88 | 3.39 | 6.86 |
| | Far | 5.38 | 0.91 | 3.59 | 7.16 |
| Mouth Height | Low | 5.25 | 0.78 | 3.72 | 6.78 |
| | Medium | 4.93 | 1.12 | 2.74 | 7.11 |
| | High | 5.39 | 0.92 | 3.59 | 7.20 |
| Smile Arc | Small | 5.06 | 1.06 | 2.98 | 7.14 |
| | Medium | 5.19 | 0.95 | 3.33 | 7.05 |
| | Big | 5.32 | 0.87 | 3.61 | 7.03 |

Regarding the main effects of feature size, we found a significant effect of mouth size ($F(2,1457) = 38.34$, $p < 0.01$), while the effect of eye size ($F(2,1457) = 1.00$, $p=0.37$), smartphone usage experience ($F(1,1457) = 1.05$, $p=0.81$), and gender ($F(1,1457) = 1.11$, $p=0.85$) were not significant. Specifically, (1) virtual faces with a mouth size of 0.27W were perceived as more trustworthy, (2) there seemed to be no significant differences between different eye sizes, (3) prior smartphone usage experience did not have a significant impact on elderly's trust in virtual agents, and (4) there was no significant effect of gender differences on elderly's trust in virtual agents (Figure 5).

Table 3: Summary of the main impacts and important interactions in ANCOVA.

| sources | df | F-statistic | p-value | MS |
|---|---|---|---|---|
| Experience (EXP) | 1 | 1.05 | p=0.31 | 0.81 |
| Gender (G) | 1 | 1.11 | p=0.29 | 0.85 |
| Eye size (ES) | 2 | 1.00 | p=0.37 | 0.77 |
| Mouth size (MS) | 2 | 38.34 | p<0.01 | 29.58 |
| Eye height (EH) | 2 | 0.56 | p=0.57 | 0.43 |
| Eye width (EW) | 2 | 13.47 | p<0.01 | 10.39 |
| Mouth height (MH) | 2 | 35.60 | p<0.01 | 27.47 |
| Smile Arc (SA) | 2 | 8.65 | p<0.01 | 6.67 |
| ES*SA | 4 | 2.99 | p<0.05 | 2.31 |
| MS*EW | 4 | 53.93 | p<0.01 | 41.61 |
| MS*MH | 4 | 11.22 | p<0.01 | 8.66 |
| MH*MS*EW | 8 | 15.64 | p<0.01 | 12.07 |
| SA*EH*EW | 8 | 2.24 | p<0.05 | 1.73 |
| EW*MH*SA | 8 | 2.11 | p<0.05 | 1.63 |

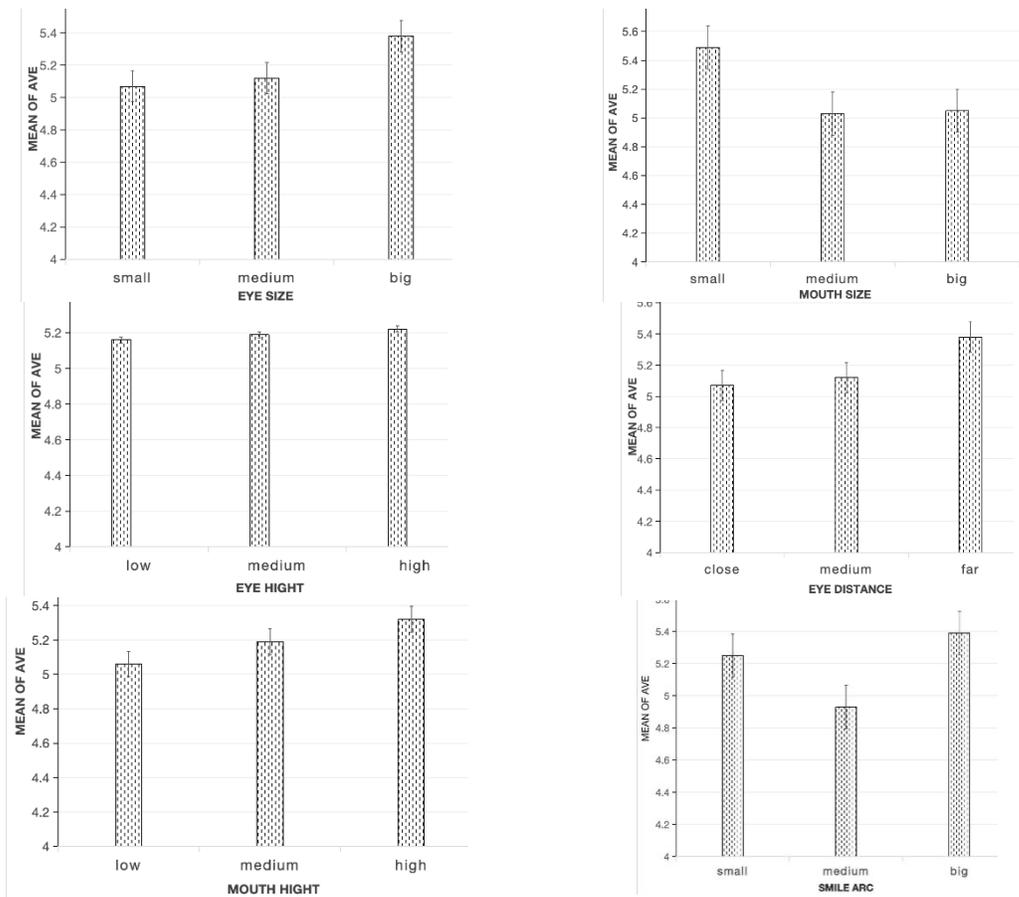

Figure 5: Single factor analysis of each characteristic item.

*4.2.2 Interaction Effects of Feature Size*

In the interaction effects of eye size with the arc of mouth smile (F(4,2914) = 2.99, p <0.05), mouth size with eye spacing (F(4,2914) = 53.93, p <0.01), and mouth size with mouth height (F(4,2914) = 11.22, p<0.01), several conclusions were drawn.

The first set of conclusions indicated that the combination of eye size at 0.25W and smile arc at 0.062H had the highest level of credibility, with a mean of 5.36(Figure 6). The second set of conclusions indicated that the combination of mouth size at 0.34W and eye spacing at 0.43W had the highest level of credibility, with a mean of 5.83. The combination of mouth size at 0.27W and eye spacing at 0.38W followed closely with a mean of 5.62. The credibility of feature combinations in the intermediate state was slightly weaker than those in the extreme states (Figure 7). The third set of conclusions indicated that the combination of mouth size at 0.27W and mouth height at 0.74H had the highest level of facial credibility, with a mean of 5.65 (Figure 8).

Similar observations were made in the three-way interaction effects of mouth size with eye spacing and mouth height (F(8,5828) = 15.64, p <0.01). The combination of mouth size at 0.27W, eye spacing at 0.41W, and mouth height at the highest level had the highest credibility. The combination of a small mouth, wide eye spacing, and low mouth height had the second-highest credibility. In contrast, the combination of a large mouth, medium eye spacing, and low mouth height had the lowest credibility. The minor mouth feature was highly recognized (Figure 9).

In conclusion, the differences in facial credibility among the three eye size proportions were insignificant. HI needed to be supported. On the other hand, mouth size significantly impacts facial credibility, with faces possessing a mouth size of 0.27W enjoying relatively higher facial credibility. H2 was confirmed.

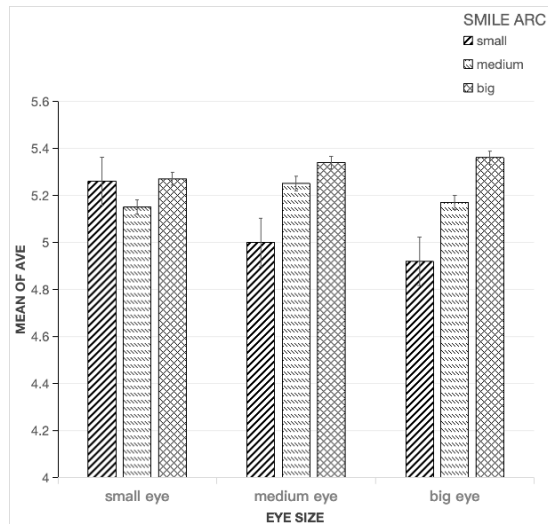

Figure 6: Bar chart of the two-way interaction results of eye size * mouth smile in facial credibility rating. Error bar represents ± 1SE.

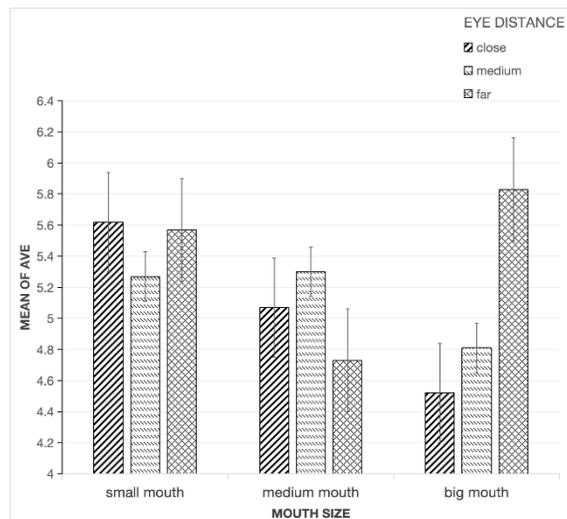

Figure 7: Bar chart of two-way interaction results of mouth size * eye spacing in facial credibility rating. Error bar represents ± 1SE.

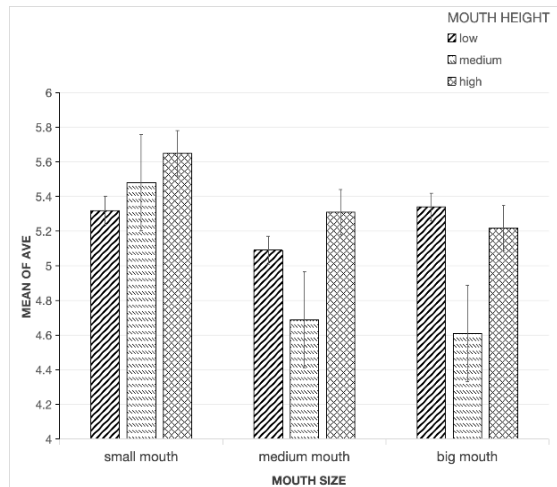

Figure 8: Bar chart of two-way interaction results of mouth size * mouth height in facial credibility rating. Error bar represents ± 1SE.

### 4.3 The effect of feature localization

#### 4.3.1 Main Effects of Feature Position

The ANCOVA results in Table 3 show strong main effects of eye spacing ($F(2,1457)=13.47$, $p <0.01$), mouth height ($F(2,1457) = 35.60$, $p <0.01$), and smile intensity ($F(2,1457)=8.65$, $p <0.01$). Specifically, (1) the highest credibility was associated with an eye spacing of 0.43W, (2) the highest credibility was associated with a mouth height of 0.74H, (3) the highest credibility was associated with a smile intensity of 0.062H, and (4) there were no significant credibility differences between different eye heights.

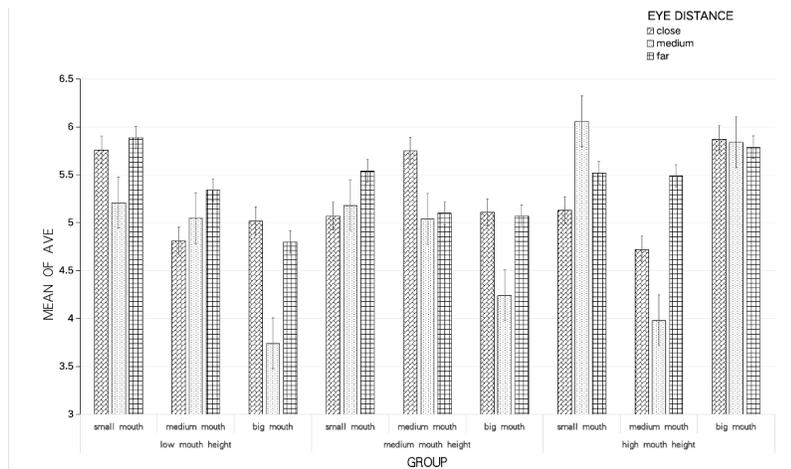

Figure 9: Bars chart representing the results of three-way interactions in facial trustworthiness rating.
(I) mouth height * eye spacing and small mouth (II) mouth height * eye spacing and medium mouth (III) mouth height * eye spacing and Da Mouth of the three-way interaction results in facial credibility rating. Error bar represents ± 1SE.

*4.3.2  Interaction Effects of Feature Position*

Additionally, we examined the interaction effects among feature positions. Regarding position interactions, the two-way interaction results in credibility ratings showed no significant effects for eye height and smile intensity (F(4,2914) = 1.14, p=0.33), eye spacing and mouth height (F(4,2914)=1.50, p 0.20), and mouth height and smile intensity (F(4,2914)=1.06, p=0.37). In terms of three-way interactions, the interaction effect of smile intensity, eye height, and eye spacing (F(8,5828)=2.24, p <0.05) suggested a trend where participants perceived higher facial credibility with increasing smile intensity, wider eye spacing, and higher eye height (Figure 10). Moreover, the interaction effect of smile intensity, eye spacing, and mouth height (F(8,5828)=2.11, p<0.05) indicated that the combination of a smile intensity of 0.062H, eye spacing of 0.43W, and mouth height of 0.74H had the highest facial credibility, while the combination of a smile intensity of 0.025H, eye spacing of 0.38W, and mouth height of 0.77H had the lowest facial credibility (Figure 11).

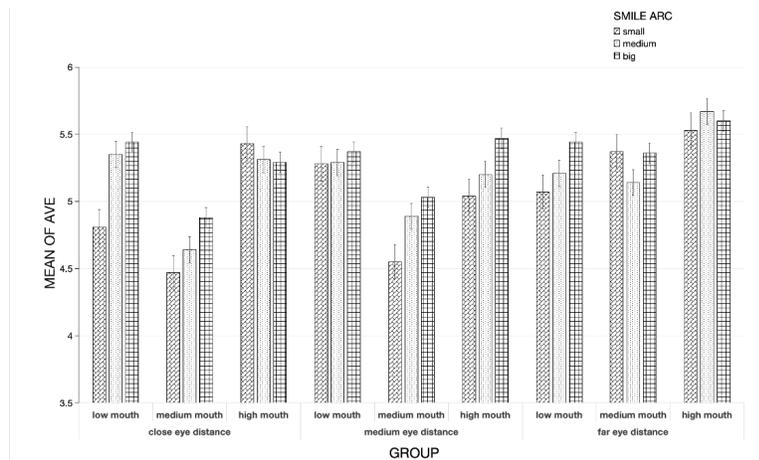

Figure 10: Bar chart showing the results of three-way interaction in facial credibility rating.
(I) eye spacing * smile degree and low mouth; (II) eye spacing * smile degree and medium high mouth;
(III) eye spacing * smile degree and high mouth. Error bar represents ± 1SE.

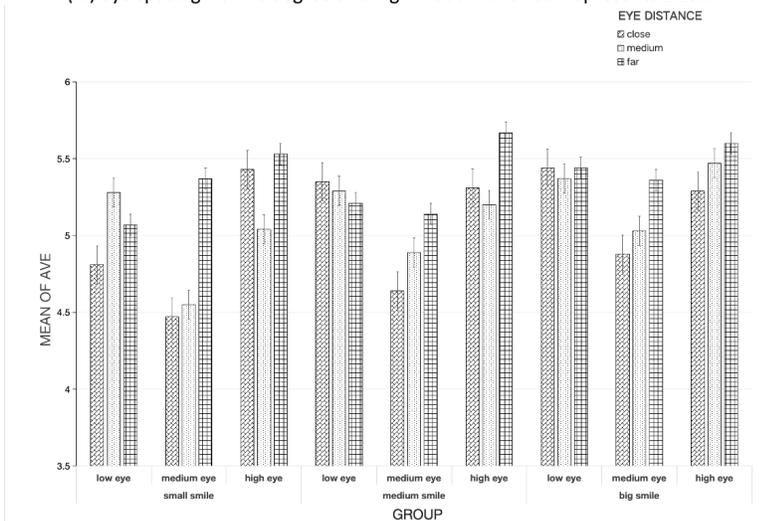

Figure 11: Bar chart showing the results of three-way interaction in facial credibility rating.
(I) smile degree * eye spacing and low eyes (II) smile degree * eye spacing and medium height eyes;
(III)smile degree * eye spacing and high eyes. Error bar represents ± 1SE.

In summary, concerning the eye region, eye height does impact facial credibility, but faces with an eye height of 0.64H enjoy relatively higher facial credibility. H3 was not supported. On the other hand, inter-eye distance significantly affects facial credibility, with faces having an inter-eye distance of 0.38W showing lower credibility compared to faces with inter-eye distances of 0.41W and 0.43W. H4 was confirmed.

Regarding the mouth region, mouth position does significantly impact facial credibility, with faces having a mouth height of 0.74H exhibiting lower credibility than faces with mouth heights of 0.77H and 0.79H. H5 was confirmed. Furthermore, the degree of smiling significantly influences facial credibility, with faces displaying a smile degree of 0.043H enjoying relatively higher facial credibility. H6 was confirmed.

### 4.4 Theoretical verification

Based on the results of the ANCOVA, we were able to identify the top three examples of virtual agent faces with the most credible and least credible facial features (Figure 12). To further validate whether these examples can truly be perceived as child faces by elderly, we recruited an additional 36 participants (mean age 66.47, SD=5.640;17 males and 19 females) from the same resource.

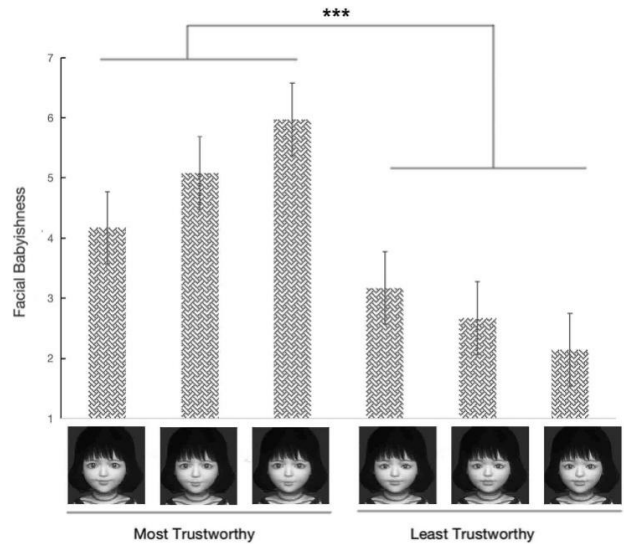

Figure 12: The Bar chart shows the results of single factor ANCOVA in the face childishness rating.
*** Significant value#0.001. The error line represents ± 1SE.

Regarding education level,13 participants reported having a primary school education or below, 14 participants reported having a secondary school education, and 9 participants reported having some education above high school.

Regarding smartphone usage experience,8 participants reported never using a smartphone,4 participants reported using one for 0-1 year (excluding 1 year),8 participants reported using one for 1-2 years (excluding 2 years), and 16 participants reported using one for more than 2 years. A one-way ANCOVA was conducted with the six examples as the between-subjects factor, smartphone usage experience as the covariate, and perceived juvenility considered as the dependent variable. Facial juvenility is measured using a 7-point Likert scale (single-item measure: 'I consider this female face to have a typical babyface appearance').

The results of the one-factor ANCOVA revealed a significant effect of the different examples ($F(5, 210)=56.5$, $p<0.001$), indicating that individuals faced with a credible virtual person may experience significantly higher levels of facial juvenility (mean 5.07; SD=1.302) compared to those faced with an untrustworthy virtual person (mean 2.66; SD 1.327)(Figure 12).

The image of the face with the highest credibility (Figure 13) and its composition structure (Figure 14) are shown below, with an eye size of 0.25W, mouth size of 0.27W, eye height of 0.64H, eye spacing of 0.41W, mouth height of 0.77H, and smile intensity of 0.043H. Consequently, we can conclude that when designing facial features of virtual agent faces for older adults, based on a facial structure with a height-to-width ratio conforming to the "baby schema" effect, the depicted face with the highest credibility can serve as a reference.

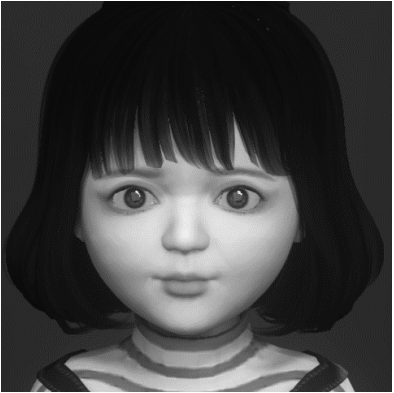

Figure 13: The face with the highest level of credibility.

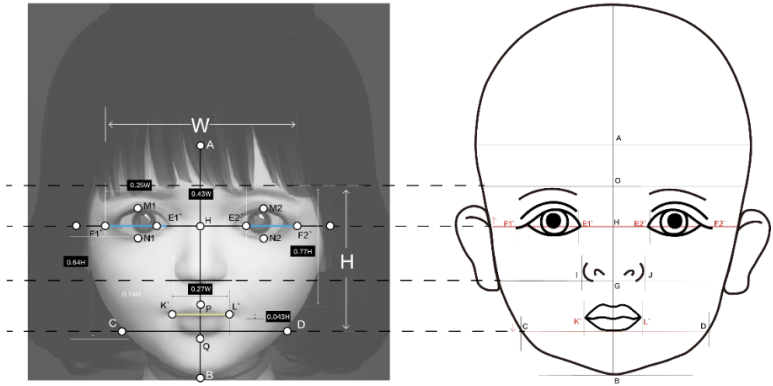

Figure 14: Facial Design Paradigm of Virtual Human Agent.

## 5 DISCUSSION

This study explored the influence of primary facial features on credibility perception among older adults. A full-factor mixed design experiment provided preliminary indicators regarding the impact of different facial feature characteristics on conveying facial credibility. The main conclusions for each feature are discussed as follows:

Feature Size: The size of the eyes did not significantly influence facial credibility. While more enormous eyes in humans and virtual agents are often associated with innocence and honesty (Yao, 2022), this feature can also evoke anxiety and seriousness in older adults, potentially undermining its perceived credibility (Marchetti, 2022). In contrast, mouth size significantly impacted facial credibility, with smaller mouths being associated with higher credibility. Notably, combining a small mouth and a high mouth height resulted in the highest credibility among all size feature combinations.

Feature Positioning: This study found that older adult participants demonstrated a superior ability to judge horizontal distance (such as eye spacing) compared to vertical distance (such as mouth height). The one-factor analysis suggests that older adults have a weaker perceptual judgment of mouth features in size and distance than eye features. This may affect their credibility judgments of faces with varying mouth characteristics. The most prominent effect was observed in the smile arc: in the vertical direction, a more significant height difference between the corners of the mouth and the lip line (philtrum) implied a more intense smile. Interestingly, moderately intense smiles were perceived as the least credible. This suggests that while older adults can discern subtle and ambiguous emotional cues, they tend to rely more on overall facial expressions, with moderate smile intensity resulting in the highest credibility when considered within the broader context of the face.

Combined features also diminished the significance of individual factors. For example, when combined with the same mouth height, a broad smile made the mouth appear lower. older adults' ability to decode emotional meaning solely from facial expressions, without the benefit of accompanying contextual cues, is weaker than that of younger adults (Riediger, 2014). As a result, less expressive facial cues may help reduce cognitive load for older adults.

Participant Insights: Despite efforts to analyze the impact of facial features on credibility perception, post-experiment interviews revealed that many older adult participants paid attention not only to the size, shape, and positioning of the eyes and mouth but also to other subtle details. Regarding the eyes, longer eyelids, and naturally upturned outer corners were considered more attractive, while downward-angled eyes gave a gentler, more pleasant impression. For the mouth, excessively thick lower lips were perceived as less genuine, and lips that turned outward were seen as less authentic. Although older adults experience declines in vision, memory, and attention compared to younger adults, their rich experience in observing faces should not be underestimated. When judging overall facial features, they pay close attention to other details, such as the height of the nasal bridge influencing the perception of eye distance, eyebrow movements when smiling, and the rounded shape of the chin, which aligns with the "baby schema" characteristics. These factors should be considered in future studies on credibility perception among older adults.

## 6 CONCLUSION

This research explored the influence of two primary facial features—eye and mouth proportions—on the perceived credibility of virtual humanoid agents to determine whether the "baby schema" effect, known for eliciting high levels of trust, can be applied to the facial design of virtual agents targeting older adults. The study contributes to the theoretical and applied exploration of human-computer interaction credibility by providing initial evidence for extending the "baby schema" effect to virtual agents designed for older adults, using a mixed experimental design to analyze the individual and interactive effects of feature sizes and positions. It further applies aging adjustments to the "baby schema" effect, proposing a paradigm for virtual humanoid facial design tailored to older adults, which incorporates detailed proportional relationships.

Although this study offers valuable insights, it also has several limitations. The research primarily focuses on the upper and lower thirds of the face, including features of the eyes and mouth, as well as specific hairstyles. However, further research into additional facial features such as cheek size, jawline curvature, and hair shape could enhance our detailed understanding of how facial features contribute to perceived trustworthiness. Additionally, although the experimental design incorporated three levels for each factor, creating 729 unique scenarios, exploring additional levels could further refine the predictive model. The participants were older adults aged between 60 and over 80 years, and future studies should investigate different subgroups within this demographic to better understand the complexities of facial credibility judgments. It is also important to note that the experimental phases lacked sufficient consideration for participants' emotional expressions and interaction environments, particularly their experiences with the "baby schema" effect in a mobile context. A more comprehensive analysis of sample biases would improve the reliability of the research. Furthermore, the "baby schema" effect might vary among older adults with different cultural backgrounds and unique life experiences, such as those living long-term with their grandchildren, which could pose challenges when generalizing these findings to other demographic groups or settings. Future research should incorporate more factors and levels to construct a more comprehensive model of facial credibility.